\documentclass[preprint,12pt]{elsarticle}

\usepackage{graphicx}
\usepackage{amssymb}
\usepackage[english]{babel}
\usepackage[utf8]{inputenc}
\usepackage[T1]{fontenc}
\usepackage{amsmath}
\usepackage{hyperref}
\usepackage{fancyhdr}
\usepackage{mathtools}
\usepackage{eso-pic}
\usepackage{microtype}
\usepackage{caption,subfigure,multicol,booktabs,array}
\usepackage{indentfirst}
\usepackage{appendix}
\usepackage{xcolor}
\usepackage{braket}
\usepackage{lineno}
\newcommand{\be}{\begin{equation}}
\newcommand{\ee}{\end{equation}}

\journal{Journal Name}

\begin{document}

\begin{frontmatter}


\title{Two Coupled Double Quantum Dots Systems as an working substance for Heat Machines}



\author[ufpb]{Jefferson Luan Diniz de Oliveira}
    \address[ufpb]{Departamento de F\'{i}sica, Universidade Federal da Para\'{i}ba\\
 Caixa Postal 5008, 58059-970, João Pessoa, PB, Brazil}
\author[ufla]{Moisés Rojas}
\author[ufla]{Cleverson Filgueiras}
\address[ufla]{Departamento de F\'{i}sica, Universidade Federal de Lavras\\
 Caixa Postal 3037, 37200-900 Lavras-MG, Brazil}

\begin{abstract}
This paper presents a conceptual design for quantum heat machines using a pair of coupled double quantum dots (DQDs), each DQD with an excess electron to interact, as an working substance. We define a compression ratio as the ratio between the Coulomb {couplings} which describes the interaction between the electrons during the isochoric processes of the quantum Otto cycle and then we analyse the arising of different regimes of operations of our thermal machine. We also show how we can achieve a {classically inconceivable Otto engine}, when considering the effects due to the {parameters related to the quantum tunneling} of a single electron between each individual DQD.   
\end{abstract}


\end{frontmatter}


\section{\Large{Introduction}}

Despite thermodynamics and quantum mechanics seems to be, at first sight, contradictory theories, early in 1959 Scovil and Schulz-DuBois \cite{Maser} demonstrated through an equivalence between a three level maser and a Carnot heat engine that it is possible to conciliate them. Since them, there are plenty of proposals for quantum heat engines and how the thermodynamic processes, or even the laws of thermodynamics \cite{bertulio}, can be defined in the microscopic world of quantum mechanics \cite{H.T.Quan}.

Beyond the quasi-equilibrium point of view of thermal machines, the pioneer work of F.L. Curzon \& B. Ahlborn \cite{Curzon} started the search for \textit{endoreversible machines}, which takes into account the time it takes for the thermodynamic processes to occur. This enables the search for the maximum power output of a heat engine, which is the most important physical quantity for practical purposes \cite{endorev,endorev2}. In the realm of quantum mechanics, the time is not a physical observable, but we highlight here the work of R. Alicki \cite{Alicki} that formally defines the thermodynamic concepts for the quantum open systems in the Markovian regime that consider a finite period for each thermodynamic cycle to be started and finished. In the last decades many proposals has been settle for quantum heat engines with different working substances, such as trapped ions \cite{Abah,trappedions,trappedions2,FILGUEIRAS2019102556}, quantum oscilators \cite{oscil,oscil2,oscil3}, Heisenberg $XX$, $XY$, $XXX$, $XXZ$ or $XYZ$ spin models with {Dzyaloshinskii–Moriya} interaction \cite{Huang_2013,He_2012,XXX,XXZ,doi:10.1142/S0217979220502124}, quantum dots \cite{QD,DQD}, etc. 

On the other hand, there is a great interest in the study of quantum dots in different aspects, like its optical and electronic properties in the production for displays \cite{display,display2} or photo-voltaic devices \cite{solarcells,solarcells2} or even in the context of quantum information processing \cite{QuantumInformation,QuantumInformation2,QuantumInformation3}. A quantum dot is a semiconductor particle and it is sometimes called an "\textit{artificial atom}" because of its similarities with a real one. The differences between them are their size (at least three orders of magnitude greater than an atom), their shape and the strength of the confining potential. Double quantum dots {(DQDs)} are, as the name suggests, two quantum dots coupled in series \cite{PhysRevLett.103.056802,doi:10.1116/1.1771679}. Then, it is straightforward to understand why they are sometimes called "\textit{artificial molecules}". The quantum dynamics and entanglement of two electrons inside the coupled {DQDs} were investigated in Refs. \cite{OLIVEIRA2015244,PhysRevB.101.075306} and the aspects related to the quantum correlations and to the decoherence were addressed in Refs. \cite{Fanchini_2010,PhysRevA.100.042309}. The Ref. \cite{Cleverson} gives us a picture of a pair of coupled {DQDs}, showing the behaviour of thermal entanglement and the correlated coherence behaves in this system as we adjust some parameters. Following this scenario, in this contribution, we implement the thermodynamic concepts of a quantum heat engine for this very same system. Each {DQD} is filled with a single electron that can tunnel (or not) between each individual island. This is possible due to quantum tunneling and Coulomb blockade effects \cite{Mesoscopic,Gorman}, which are observed in very small devices, as in our case: we consider quantum dots separated by tunnel junctions that act as an insulating barrier. The quantum tunneling has a central role in our findings since it shows to be a parameter that may changes the operational mode of a heat machine we are proposing. Although the paradigm in this field is the investigation of finite time cycles in open quantum systems, at this stage we focus in the quasi-static process so the the system will be described at the end of an isochoric strokes by a Gibbs state, which means that it is in a thermal equilibrium with a heat bath. Nevertheless, we will present important insights which we believe are going to survive in this context but need to be clarified. Here, it is shown how {we can find nonclassical results based in the control of the effects} due to a quantum tunnelling of a single electron between each individual {DQD}. The performance of both the engine and the refrigerator can be improved and the operation mode of the machine can be yet altered, yielding the possibility of work extraction for an incompressible working substance. 

This paper is organized as follows: in Section 2 we present a general overview of the system and how we are going to accost it. In Section 3 we discuss the processes involving the quantum Otto cycle and then calculate both the total work done and the engine efficiency. In Section 4 we discuss the results and analyse minutely the emergence of different regions of regime for the quantum heat engine. In Section 5 we present the concluding remarks that summarize our results.

\section{A pair of double quantum dots as an working substance}
\begin{figure}[ht!]
	\centering
	\includegraphics[height=5cm,width=6cm]{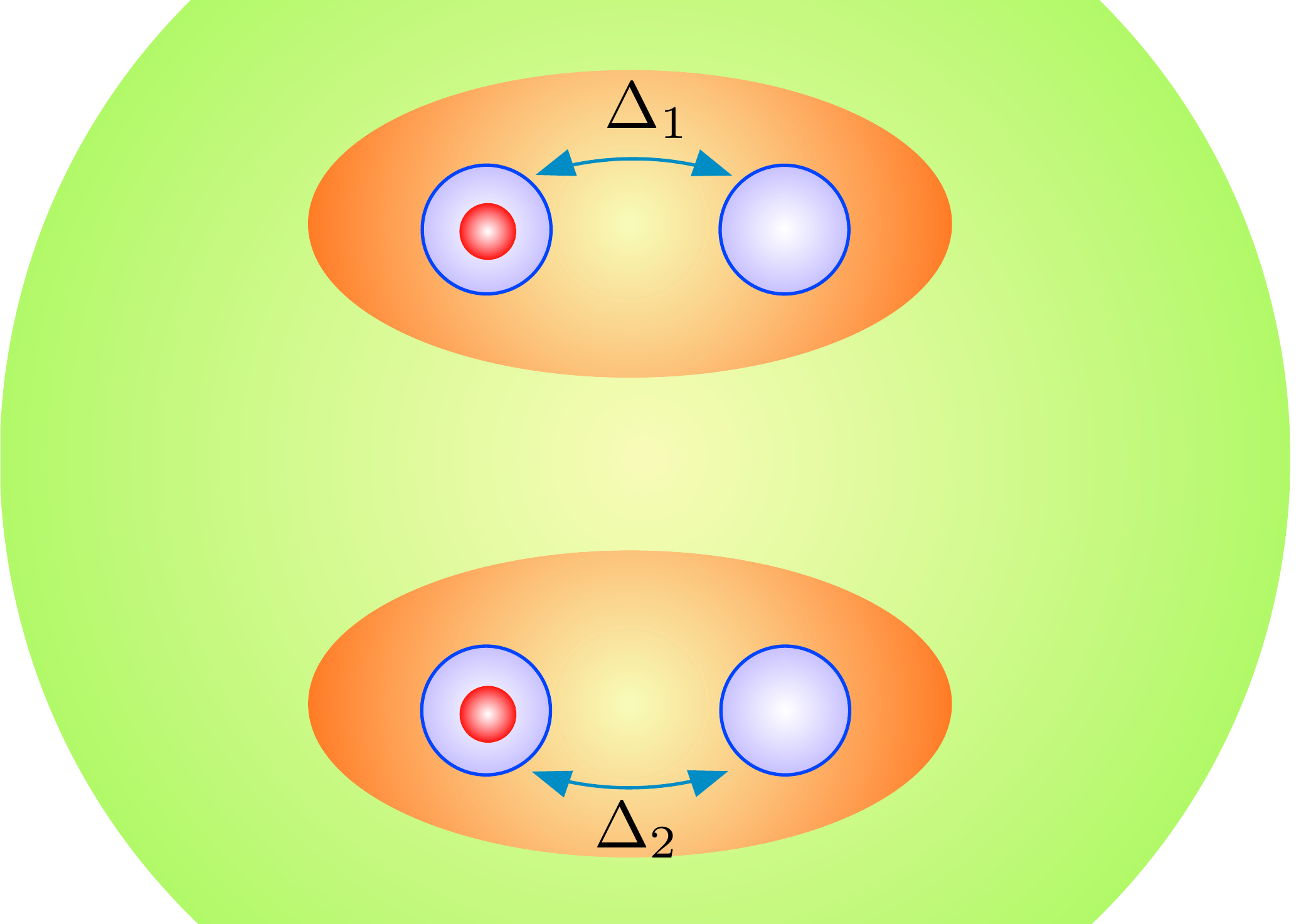}
	\quad
	\includegraphics[height=5cm,width=7cm]{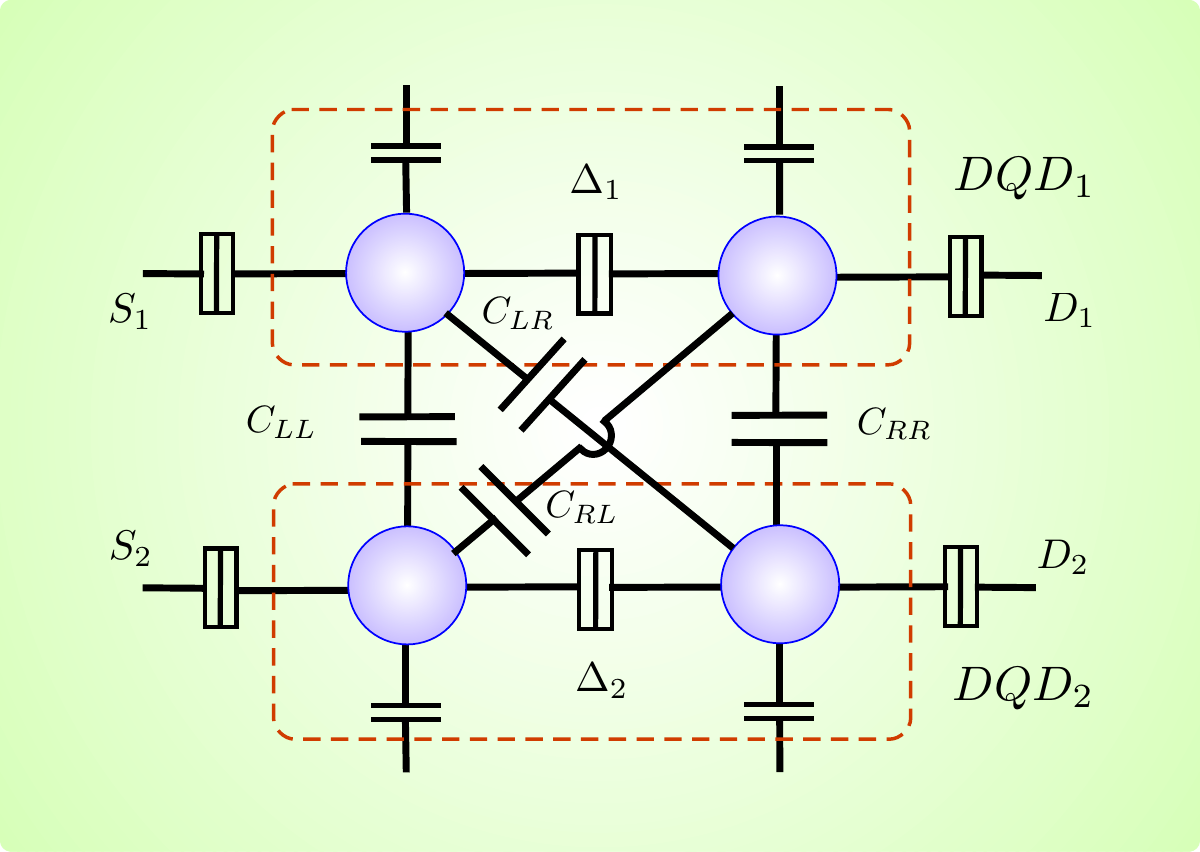}
	\caption{\small {The left picture shows} a schematic representation of the physical model with two coupled DQDs. The purple spheres represent the quantum dots, the electrons are represented by the smaller red spheres inside the quantum dots and $\Delta_{1,2}$ stands for the tunneling coupling of the DQD$_{1,2}$. {The right picture is the equivalent circuit diagram of the device. Tunnel couplings are represented by two bars as in $\Delta_1$ and $\Delta_2$, $S_{1,2}$ and $D_{1,2}$ are, respectively, the source and drain of the DQD$_{1,2}$, and the capacitors $C_{LL}$, $C_{RR}$, $C_{LR}$ and $C_{RL}$ connect the DQDs.}}\label{fig0}
\end{figure}

{The proposed quantum system as an working substance for quantum heat engines can be described by the two-qubit device that is fabricated in a standard AlGaAs/GaAs heterostructure with two-dimensional electron gas (2DEG) by using electron beam lithography. The device consists of a series  of Schottky  gates\footnote{{The Schottky gates are not illustrated in Figure \ref{fig0} due to the high number of gates usually implemented experimentally.}} and the electron configurations in both DQDs are controlled by tuning the voltages applied on the gates {(see Figure \ref{fig0}).} On the other hand, the two double quantum dots are capacitively coupled, this capacitance model describing the Coulomb interactions between the two DQDs \cite{PhysRevLett.103.056802,FUJISAWA2011730,PhysRevLett.103.016805}.} {This capacitance coupling is controlled by the capacitors $C_{LL}$, $C_{RR}$, $C_{LR}$ and $C_{RL}$, where no transition for the electron to leave the DQD$_1$ into DQD$_2$ is allowed, and vice-versa. The tunnel junctions placed after the source $S_{1,2}$ and before the drain $D_{1,2}$ are adjusted in such a way to make the system in the Coulomb blockade regime, enabling a single electron to be confined in each DQD}. The charge of the electron in each {DQD} build up the qubits, which are described by the two possible states for the location of each electron, the left dot ($\ket{L}$) and the right dot ($\ket{R}$), {where the electron can tunnel from left to right and from right to left}. The Hamiltonian of such a system is given by
\be\label{eq1}
H=\Delta_1\sigma_{1}^x+\Delta_2\sigma_{2}^x+V(\sigma_{1}^z \otimes \sigma_{2}^z),
\ee
where $\Delta_1$ and $\Delta_2$ are the strength of the tunneling coupling between each pair of quantum dots, $V$ is the interaction Coulomb {coupling} between the excess electrons and $\sigma_{1(2)}^{x,y,z}$ are the Pauli matrices. {The tunneling coupling parameters $\Delta_1$ and $\Delta_2$ are controlled by the gate voltages (Schottky gates) and the interaction coupling $V$ is controlled by the capacitors connecting both DQDs}. A more general version of the Hamiltonian of this system takes into account an extra term for the energy differences between the uncoupled charged states $\ket{L}$ and $\ket{R}$, but the difficulties behind solving its equations are not in the scope of our aim in this paper. This way, we consider the simplest case where all of our quantum dots have the same energy available for the electron to occupy.

\begin{figure}[ht!]
	\centering
	\includegraphics[height=7cm,width=8cm]{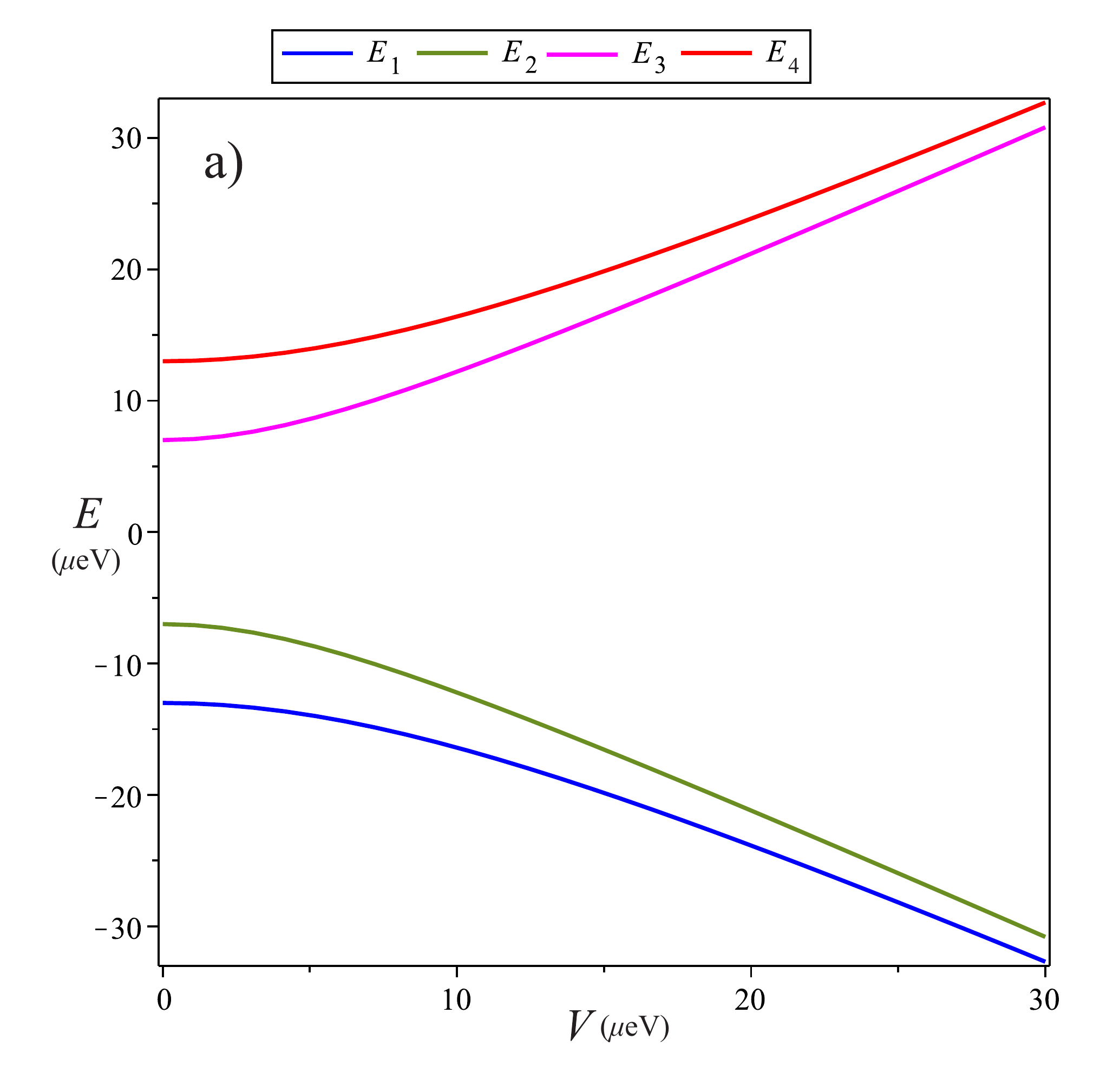}
	\quad
	\includegraphics[height=7cm,width=8cm]{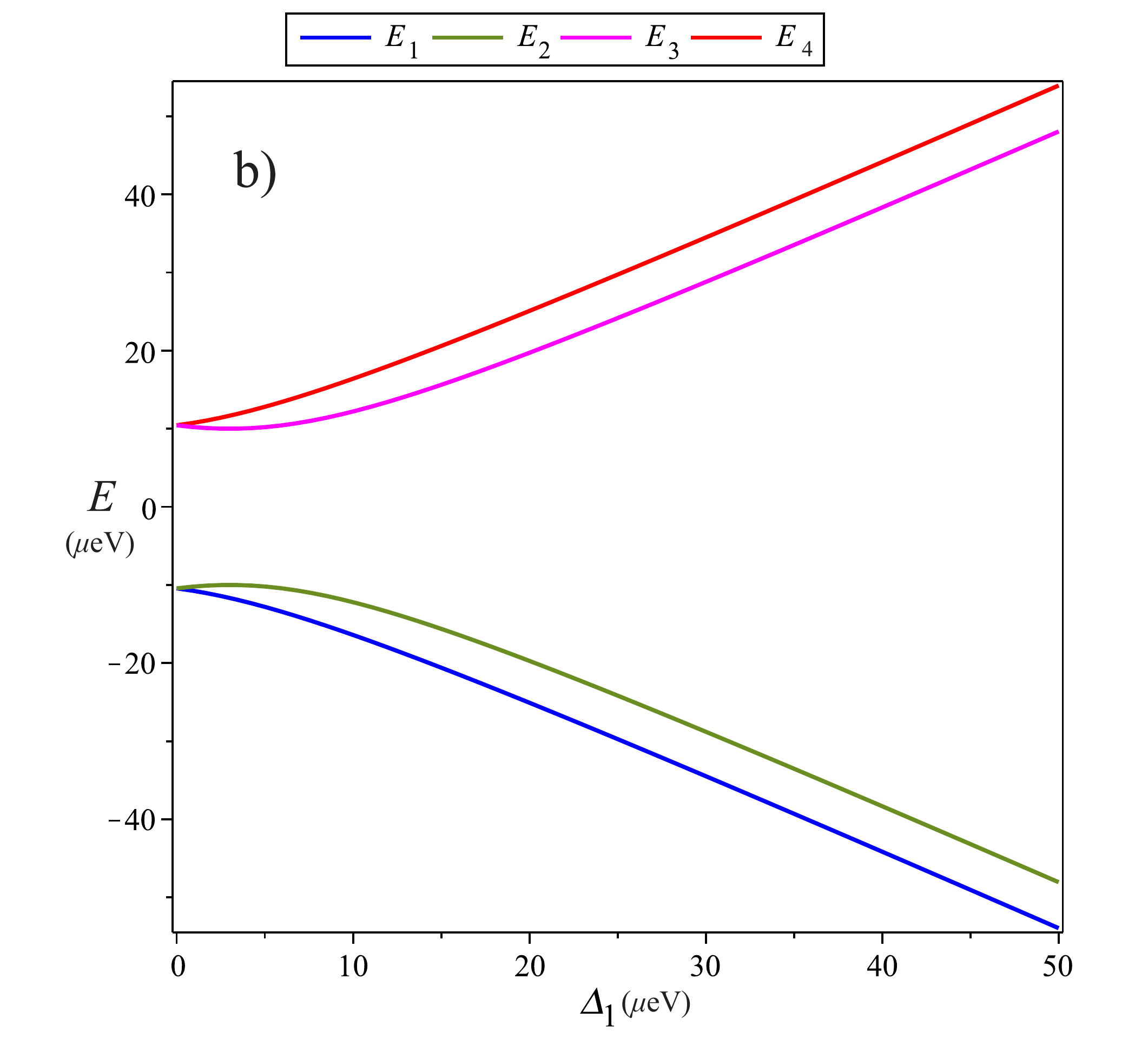}
	\caption{\small In a), it is depicted the energy levels in terms of the interaction {coupling} $V$ between the two {DQDs} for the same fixed tunneling parameters $\Delta_1=10\mu \text{eV}$ and $\Delta_2=3\mu \text{eV}$ . In b), we have the plot of the energy levels against the tunneling parameter $\Delta_1$ for fixed $\Delta_2=3\mu \text{eV}$ and $V=10\mu \text{eV}$. Notice that there is a squeezing of the energy gaps as we either increase the {interaction coupling} or decrease the tunneling parameter.}\label{fig1}
\end{figure}

Solving the eigenvalue equations for the Hamiltonian (\ref{eq1}), we obtain the following eigenstates (see Ref. \cite{Cleverson})

\begin{equation*}			 	\ket{\psi_1}=\alpha_-[A_-(-\ket{LL}+\ket{RR})+n_-(\ket{LR}-\ket{RL})],
\end{equation*}
\begin{equation*}
	\ket{\psi_2}=\alpha_-[n_-(-\ket{LL}+\ket{RR})+A_-(-\ket{LR}+\ket{RL})],
\end{equation*}
\begin{equation*}
	\ket{\psi_3}=\alpha_+[A_+(\ket{LL}+\ket{RR})+n_+(\ket{LR}+\ket{RL})],
\end{equation*}
\be
	\ket{\psi_4}=\alpha_+[n_+(\ket{LL}+\ket{RR})-A_+(\ket{LR}+\ket{RL})],
\ee
where $\alpha_\pm=\frac{1}{\sqrt{2}\sqrt{(n_\pm)^2+A_{\pm}^2}}$, $A_\pm=V+\sqrt{(n_\pm)^2+V^2}$ and $n_\pm=\Delta_1\pm\Delta_2$, with the following eigenenergies

\be
E_{1}=-\sqrt{(n_-)^2+V^2},
\ee

\be
E_{2}=-\sqrt{(n_+)^2+V^2},
\ee

\be
E_{3}=\sqrt{(n_-)^2+V^2},
\ee

\be
E_{4}=\sqrt{(n_+)^2+V^2}.
\ee
An important result that we will later need is how the eigenenergies change as either the Coulomb {coupling} or one of the tunneling parameters increases (it does not matter which of them we take since the Hamiltonian is symmetric as we change $\Delta_{1}\leftrightarrow \Delta_2$). From the Fig. \ref{fig1} we see that the energy levels are compressed pairwise and, at the same time, the ground and the first excited state are separated from the second and third excited state as the interaction {coupling} is raised. On the other hand, the energy levels are pairwise detached. As we increase the tunneling parameter, they are shifted apart from each other more rapidly than the energy levels separation observed inside each pair separately. Thus the energy scale variation is not uniform. This way, {as a first approach}, we will approximate our system by a two level one with energies $E_1$ and $E_2$.
	
{Despite the values of the tunneling parameters, $\Delta_1$ and $\Delta_2$, are mostly predetermined in the fabrication of the device, we still can modify them experimentally  \cite{PhysRevLett.103.056802}.}


Here we {refer the system as in a Gibbs state when it is in a thermal equilibrium with a heat bath, which means that its density matrix is given by $\rho(T)=\frac{exp(-\beta H)}{Z}$, where $Z$ is the partition function, $\beta=\frac{1}{kT}$, $k$ is the Boltzmann constant and $T$ is the temperature of the heat bath.}

\section{The Quantum Otto Engine Cycle}
\begin{figure}[ht!]
	\centering
	\includegraphics[height=9cm,width=13cm]{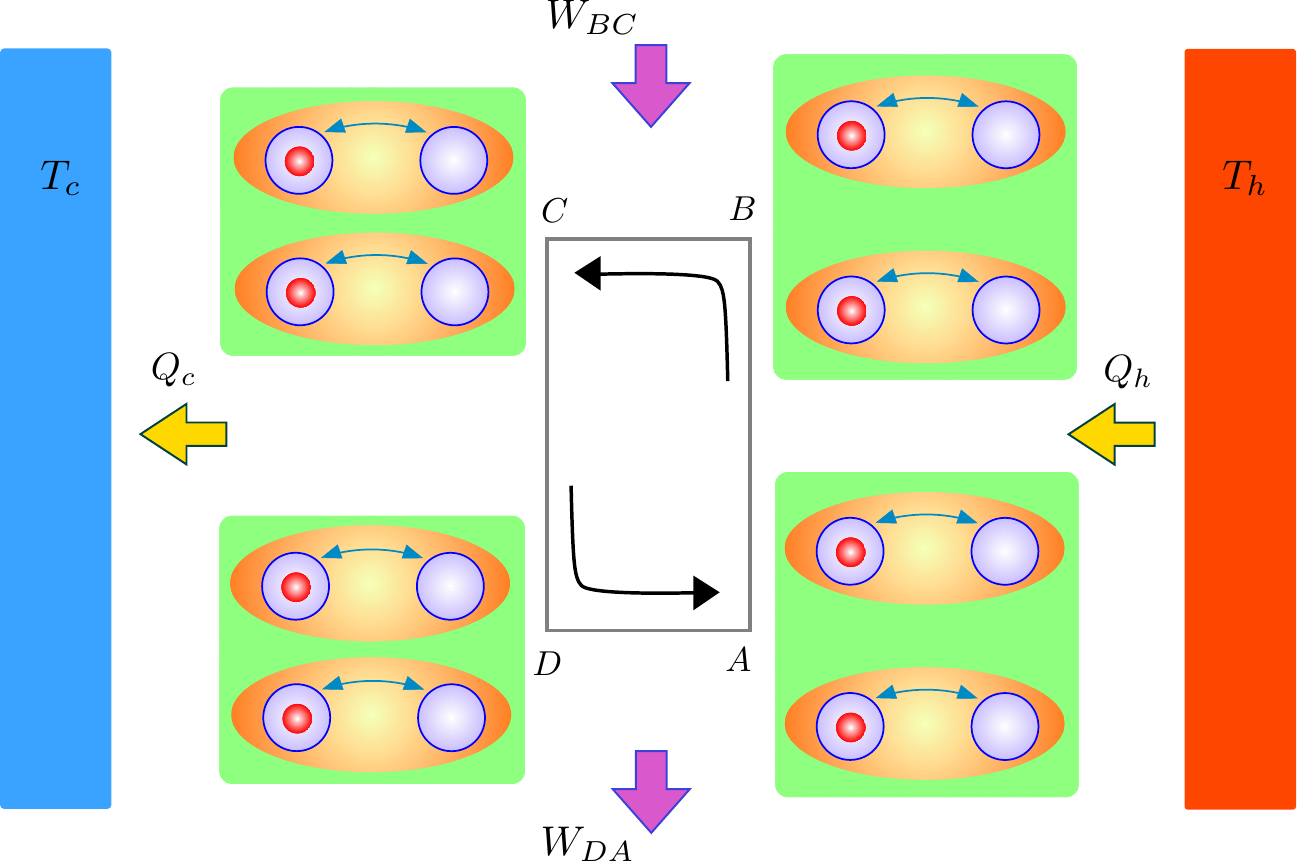}
	\caption{\small A schematic representation of an Otto engine using a pair of coupled {DQDs} as a working substance: the incoming heat from the hot bath, $Q_h$, is transformed into extracted work. The engine cycle consists of two adiabatic strokes ({$B\rightarrow C$ and $D\rightarrow A$}) where it is decoupled from the thermal baths, and two isochoric strokes ({$A\rightarrow B$ and $C\rightarrow D$}) where the engine is coupled to two thermal baths at temperatures $T_h$ and $T_c$, with $T_h>T_c$.}\label{figa}
\end{figure}

In this section we describe the quasi-static quantum Otto engine cycle, which operates in four strokes: two quantum isochoric processes and two quantum adiabatic processes (see Fig. \ref{figa}).

The cycle starts with a quantum isochoric process {($A\rightarrow B$)}: the working substance, {with interaction {coupling} $V_h$ and tunneling parameters $\Delta_1^h$ and $\Delta_2^h$}, is put in contact with the hot reservoir at temperature $T_h$ until they reach a thermal equilibrium and a total heat $Q_h>0$ is transferred to the system at the end of the process. {The parameters $V$, $\Delta_1$ and $\Delta_2$ that regulate the eigenenergies are controlled externally, what makes this process easier to realize in an experiment}. At the end of the process we will have, {for the energy eigenstate basis $\{\ket{\psi_n}\}$}, the following density matrix
\be\label{eq2}
\rho_h=exp(-H_h/kT_h)/Z_h=\sum_n p_{n}^h\ket{\psi_n}\bra{\psi_n},
\ee
with 
\be\label{eq3}
H_h=\sum_nE_n^h\ket{\psi_n}\bra{\psi_n},
\ee
\be\label{eq4}
p_n^h=exp(-E_n^h/kT_h)/Z_h,
\ee
\be\label{eq5}
Z_h=\sum_{n}exp(-E_n^h/kT_h),
\ee
where $Z_h$ is the partition function, $p_n^h$ is the occupation probabilities of each eigenstate and $H_h$ is the Hamiltonian when the system is in contact with the hot heat bath.

Next, we have a quantum adiabatic expansion {(B$\rightarrow C$). In this process} no heat is exchanged between the system and the environment. The working substance is totally isolated from the environment and therefore, the Eq. (\ref{eq4}) is no longer valid during the process because there is no thermal equilibrium with it. Thus we can increase the interaction {coupling} from $V_h$ to $V_c>V_h$ {by adjusting the voltages in the capacitors mentioned in Figure \ref{fig0}} and still keep the occupation probabilities $p_n^h$ constant until the end of the process. Moreover, {in parallel to this}, the tunneling parameters may be tuned from $\Delta_1^h$($\Delta_2^h$) to $\Delta_1^c$($\Delta_2^c$) by controlling the gate voltages in each {DQD}. Thereby the energies increases from $E_n^h$ to $E_n^c$, so the Hamiltonian will be given by $H_c=\sum_nE_n^c\ket{\psi_n}\bra{\psi_n}$ and some work {$W_{BC}>0$} is extracted from the system. 

In the next stroke we have another quantum isochoric process {(C$\rightarrow D$)}. We put the working substance in contact with the cold reservoir at temperature $T_c$, waiting enough time for the thermalization to occur. A total heat $Q_c<0$ is transferred to the cold reservoir at the end of the process and, since heat is exchanged, the occupation probabilities change from $p_{n}^h=exp(-E_n^h/kT_h)/Z_h$ to $p_{n}^c=exp(-E_n^c/kT_c)/Z_c$, {with $Z_c=\sum_{n}exp(-E_n^c/kT_c)$}. The density matrix will be given by $\rho_c=exp(-H_c/kT_c)/Z_c=\sum_n p_{n}^c\ket{\psi_n}\bra{\psi_n}$ and we keep the energies $E_n^c$ fixed. 

Finally, we close the cycle with a quantum adiabatic compression {(D$\rightarrow A$)}. {At this point, we adjust the voltages in the capacitors again, causing the interaction {coupling} to change from $V_c$ to $V_h$. Apart from that}, the tunneling parameters {may be tuned} from $\Delta_1^c$($\Delta_2^c$) to $\Delta_1^h$($\Delta_2^h$) and, consequently, the energies from $E_n^c$ to {$E_n^h$}. The Hamiltonian of the system {in the end of the process} is then $H_h=\sum_nE_n^h\ket{\psi_n}\bra{\psi_n}$, the occupation probabilities $p_n^h$ are kept unchanged and some work {$W_{DA}<0$} is done on the working substance.

As stated in Ref. \cite{Cleverson} we can say that when the working substance is in contact with the hot reservoir, it is in a weakly correlated state and, as we increase the {interaction coupling} and decrease the temperature, the system starts to be more correlated.

The quantum version of the first law of thermodynamics in the quasi-static limit allow us to calculate the total heat exchanged during the isochoric processes \cite{H.T.Quan}, that is,
\be\label{eq6}
Q_h=\sum_{n}E_n^h(p_n^h-p_n^c),
\ee
and
\be\label{eq7}
Q_c=\sum_{n}E_n^c(p_n^c-p_n^h),
\ee
where $Q>0$ ($Q<0$) means that heat is absorbed (released) from (to) the heat reservoirs, respectively. Therefore, the total work $W$ produced by the heat engine in the adiabatics is, by energy conservation, the excess heat
\be\label{eq8}
W=Q_h+Q_c=\sum_{n}(E_n^h-E_n^c)(p_n^h-p_n^c).
\ee
With that in hands, we can finally have the efficiency of our heat engine, which is calculated by $\eta\equiv W/Q_h$.

The description of the refrigerator cycle is totally analogous to the processes of the heat engine discussed early, except for the direction of operation of the cycle, which is reversed. This means that we will have a heat released $Q_c>0$ from the cold heat bath and absorbed $Q_h<0$ by the hot heat bath to the working substance and, consequently, it will be necessary an external work $W=Q_h+Q_c<0$ for the cycle to operate. The coefficient of performance (COP) $\varepsilon$ measures the efficiency of the refrigerator, which is defined as the modulus of the ratio of heat released from the cold heat bath and the total work done in the cycle, that is, $\varepsilon\equiv Q_c/W$. 

With those definitions in hand, in the next section we investigate in detail the work, the efficiency $\eta$ and the COP $\varepsilon$. 
{Although this work is theoretical, a possible implementation of the thermal machine with double quantum dots is to consider the recent concept of  particle-exchange heat engines, which uses energy filtering to control a thermally driven particle flow between two heat reservoirs. As they do not require moving parts and can be realized in solid-state materials, they are suitable for low-power applications and miniaturization \cite{QD}}.
\newpage
\section{Results and Discussion}

{In order to plot the next graphics, we first define the compression ratio $r$ as the ratio between the maximum and the minimum interaction coupling values in the cycle, $r=V_c/V_h$. Recalling that this interaction is due to the Coulomb interaction potential, increasing the interaction coupling is equivalent to increase the Coulomb potential and, since the Coulomb potential is inversely proportional to the distances between the electrons, this is also equivalent to decrease the distance between these electrons. Thus, the compression ratio is effectively measuring how our system is being compressed as a whole (causing the approximation of the electrons), despite we are never actually doing this, since we are only controlling voltages in the capacitors. This is exactly what we expect for a compression ratio, it measures how much our working substance is being compressed during the cycle. From this point on we restrict ourselves to the case where the tunneling parameters $\Delta_1$ and $\Delta_2$ are the same for the whole cycle and we will no longer worry about the upper index.}

\begin{figure}[ht!]
	\centering
	\includegraphics[height=8cm,width=10cm]{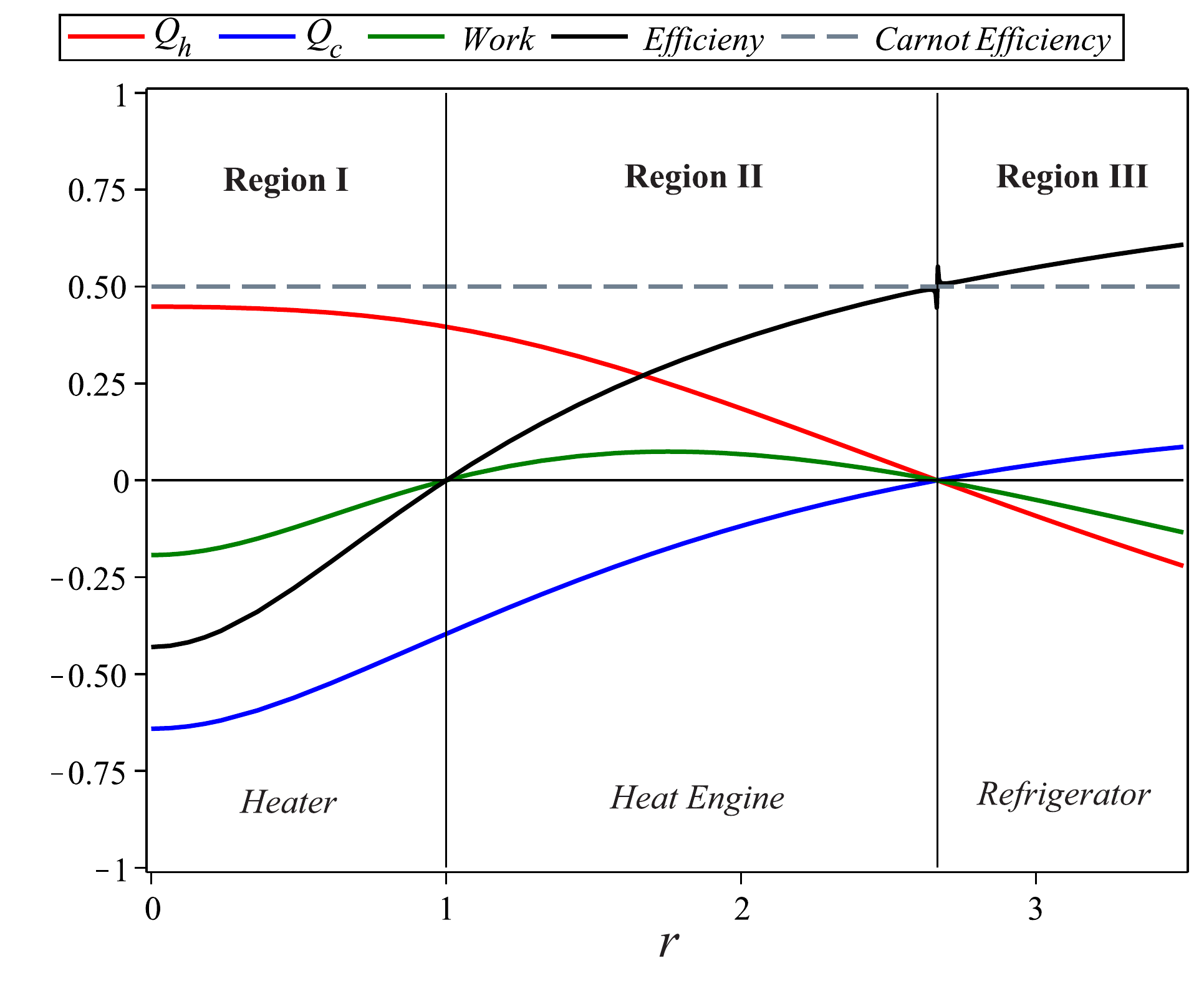}
	\caption{\small As illustrated, we have the heat exchanges of the working substance with the hot and cold reservoirs ($Q_h$ and $Q_c$, respectively), the work $W$ done, the efficiency $\eta$ and the Carnot efficiency $\eta_{\small C}$ of the heat engine against the compression ratio $r$. Heat transfer between the hot and cold reservoir. A sign inversion on the flow of the heat happens as we increase the compression ratio $r$. For this plot, we have chosen the values {$V_h=10\mu \text{eV}, \Delta_1=10\mu \text{eV},\Delta_2=3\mu \text{eV}, T_h=2\mu \text{eV},T_c=1\mu \text{eV}$.} {The values of $Q_h$, $Q_c$ and $W$ are given in units of $\mu \text{eV}$}.}\label{fig4}
\end{figure}

{In the Fig. \ref{fig4} the heat exchanged $Q_h$ ($Q_c$) with the hot (cold) reservoir, the work done $W$, the efficiency $\eta$ and the Carnot efficiency $\eta_c$ are given in terms of the compression ratio $r$, where we have set\footnote{The choice of the values for the tunneling parameters are not random: for either $\Delta_1 \approx \Delta_2$ or $\Delta_2 \gg \Delta_1$ (or even $\Delta_1 \gg \Delta_2$) {the work achieved is minimum, see Appendix A. See also Appendix B for an explanation on how these values for the reservoirs temperatures enables the two-level approximation.}} $\Delta_1=10\mu \text{eV},\Delta_2=3\mu \text{eV}$ and $T_h=2\mu \text{eV}, T_c=1\mu \text{eV}$ (we normalize the Boltzmann constant $k=1$ in this whole paper). It is clear from the Fig. \ref{fig4} that we need $r>1$ ($V_c>V_h$) to achieve positive work, which is reasonable because in this regime the energy gaps are squeezed (see Fig. \ref{fig1}) when the system is in contact with the cold reservoir and they are expanded when the system is in contact with the hot reservoir \cite{Thermodynamics}. Note that we can not increase the compression ratio $r$ indefinitely since the positive work condition is lost. As we increase the value of $r$, it comes to a point where there is no heat transfer even when the system is in contact with the hot and the cold reservoir. Furthermore, after this point, the signs of the heats exchanged are inverted and the system starts to withdraw heat from the cold reservoir and transfer heat into the hot reservoir. In other words, the system starts behaving as a refrigerator at cost of some work.} 

{To summarize}, we can see the appearing of three different regions of operation for the heat engine. In the region I the engine requires a negative work $W<0$ to extract heat from the hot to the cold reservoir, i.e, the machine operates as a heater. In the region II we have a positive work, $W>0$, which means that the it acts as a heat engine producing useful work. Finally, the region III correspond to a refrigerator as we have already discussed early, because we have an inversion on the flow of the heat for some negative work $W<0$. These results show that we can pass through the different regimes by simply increasing a single parameter, the interaction {coupling} $V_c$ (which causes the change in $r$). Another way to interpret this inversion {of heat fluxes} is by observing that as we increase {the value of the Coulomb coupling} the system gets more and more strongly correlated so that there is a value of concurrence that is critical where the signs of the transferred heats are exchanged \cite{Cleverson}.
\begin{figure}[ht!]
	\centering
	\includegraphics[height=8.5cm,width=10cm]{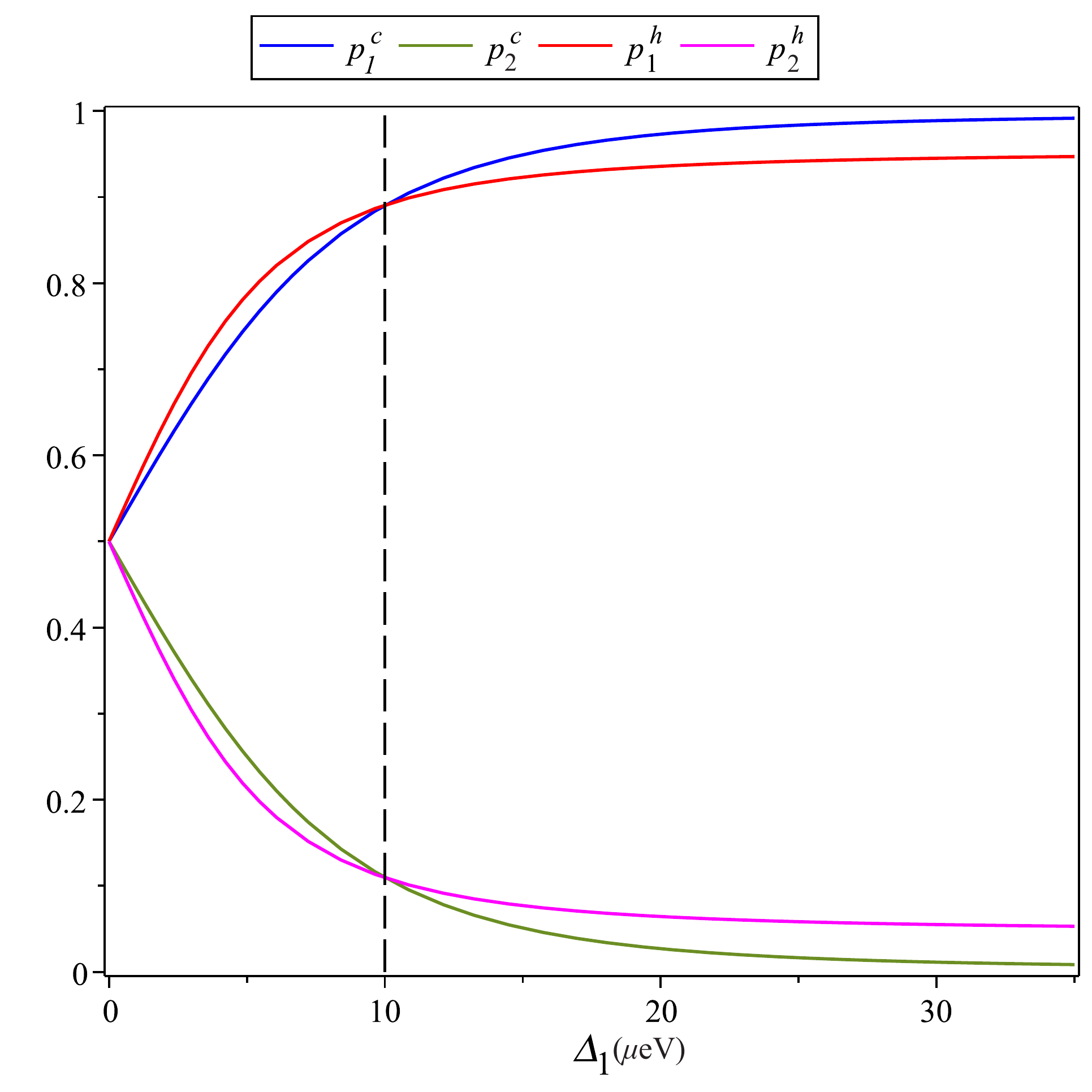}
	\caption{\small The occupation probabilities curves for the ground and first excited state in the two situations are shown, when the system is in contact with the hot heat bath and the cold heat bath. The probabilities curves for higher excited states are omitted because their values are close to zero {for this values that we take, to know, $V_h=10\mu \text{eV}, \Delta_2=3\mu \text{eV}, T_h=2\mu \text{eV},T_c=1\mu \text{eV}$ and $r=2.67$, which is approximately the value of the compression ratio where the inversion of the heat fluxes occurs}.}\label{fig5}
\end{figure}

Notice that the two points where the green curve of the work done intersects the $r$-axis in the Fig. \ref{fig4} have different meanings: the first one has to do with the equality in modulus of the heat $Q_h$ and $Q_c$ (see Eq. \ref{eq8}), i.e, all of the heat absorbed from the hot reservoir is released to the cold reservoir, and the second one has to do with the totally interruption of the heat transferred to both reservoirs. This interruption can be explained by means of the occupation probabilities $p_{n}^h=exp(-E_n^h/kT_h)/Z_h$ and $p_{n}^c=exp(-E_n^c/kT_c)/Z_c$, where we can see in the Fig. \ref{fig5} that there is a point where the occupation probabilities curves for the hot and cold heat baths intersects, which means that in this regime there is no change on the occupation probabilities of the system when it passes from the the hot heat bath to the cold heat bath, {what causes the interruption of the heat flux.}

\begin{figure}[ht!]
	\centering
	\includegraphics[{height=7cm,width=8cm}]{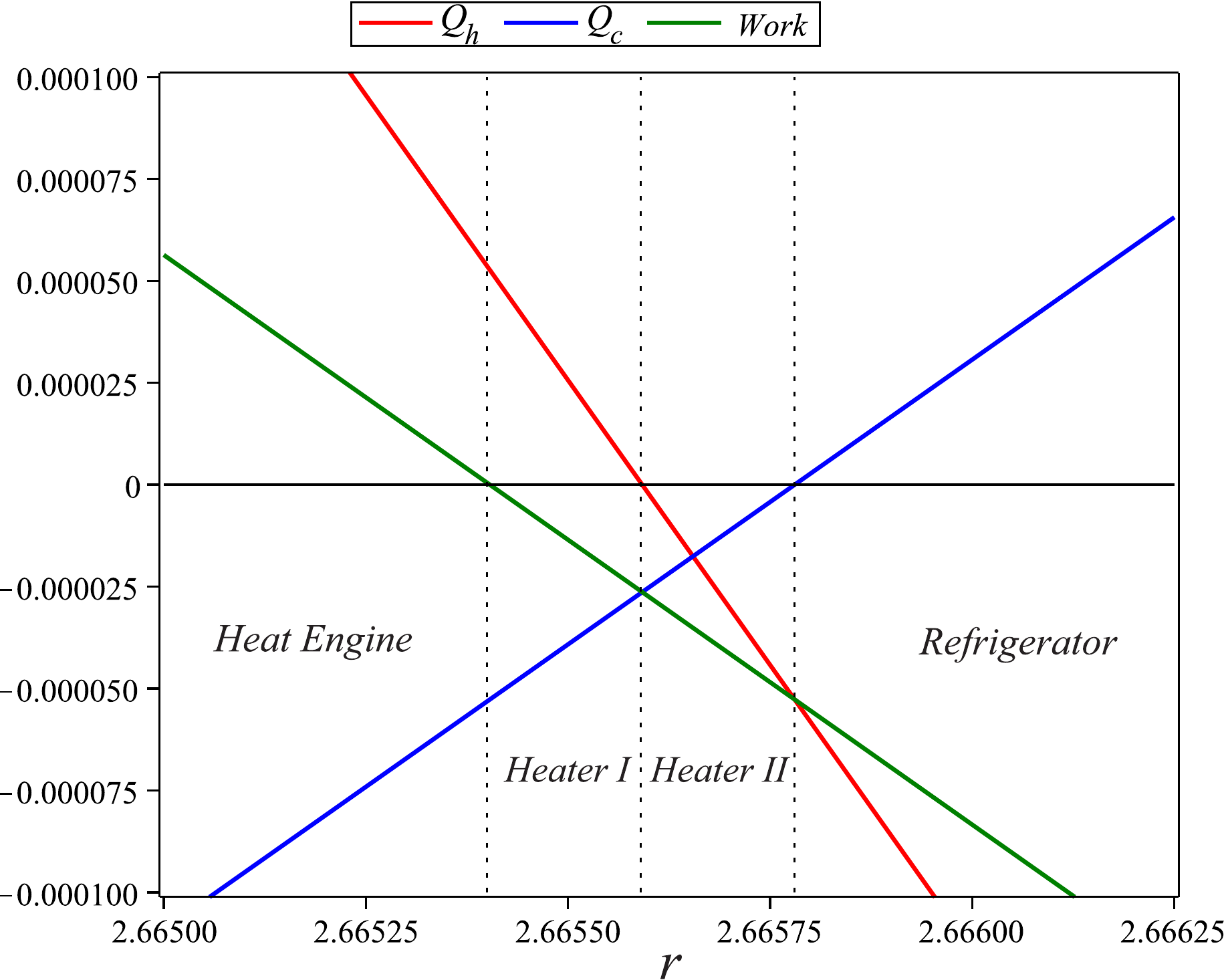}
	\caption{\small {As illustrated}, we have the work done W (green), the heat absorbed $Q_h$ (red) and released $Q_c$ (blue) against the compression ratio $r$: the observed behaviour remind us something like a phase transition. We keep the same values, as usual, {$V_h=10\mu \text{eV},\Delta_1=10\mu \text{eV},\Delta_2=3\mu \text{eV}, T_h=2\mu \text{eV},T_c=1\mu \text{eV}$.} {The values of $Q_h$, $Q_c$ and $W$ are given in units of $\mu \text{eV}$.}}\label{fig6} 
\end{figure}

After zooming the Fig. \ref{fig4}, we can extract an additional information about the "\textit{point}" that causes the divergence on the efficiency plot. First of all, as we can see from the Fig. \ref{fig6}, there is no such a thing as a point that simultaneously invert the signs of the heat exchanged $Q_h$ and $Q_c$. Before the efficiency explodes, the work tends to zero, and so the efficiency, where over again we have the heat pump regime. After an almost infinitesimal increase on $r$, there is an explosive increase on efficiency	due to the interruption on the heat flow $Q_h$ (remember that $\eta=W/Q_h$). After this point, the system does not immediately starts behaving as a refrigerator, we have a tiny region where the machine consumes work and exhaust heat for both reservoirs (we use here the notations of the Ref. \cite{PhysRevE.102.052131} for the two different kinds of heat pump, to know, \textit{heater I} for the usual heater and \textit{heater II} for the machine that heats both reservoirs, see also Ref. \cite{Chand}). Only after that, we will have a positive heat flow from the cold reservoir to the system turning the machine into a refrigerator.
\subsection{The refrigerator regime}

\begin{figure}[ht!]
	\centering
	\includegraphics[height=7cm,width=7cm]{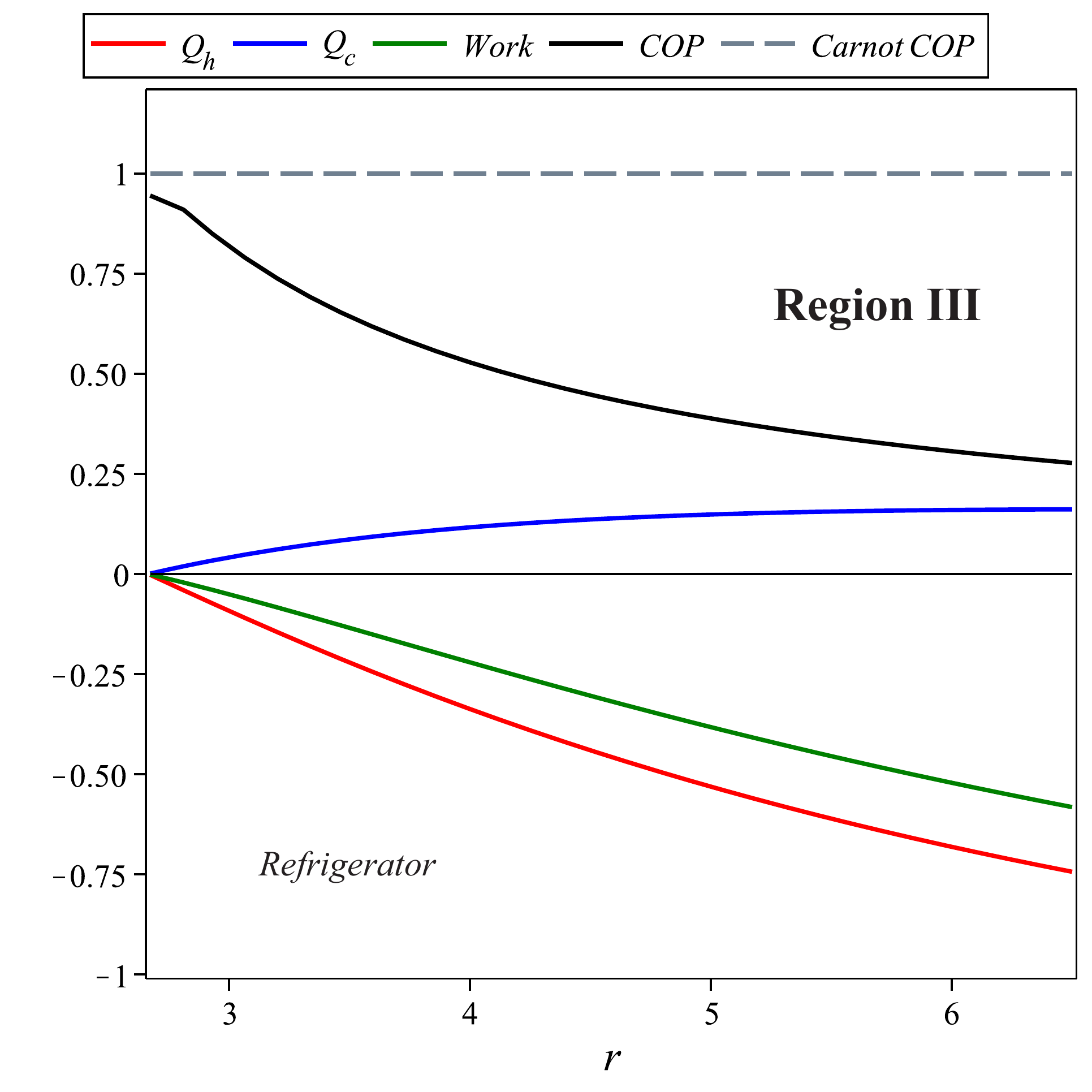}
	\caption{\small As illustrated, we have the heat exchanges of the working substance with the hot and cold reservoirs ($Q_h$ and $Q_c$, respectively), the work done and the COP $\varepsilon$ of the refrigerator against the compression ratio $r$. To plot this graph we have chosen the values {$V_h=10\mu \text{eV}, \Delta_1=10\mu \text{eV},\Delta_2=3\mu \text{eV}, T_h=2\mu \text{eV},T_c=1\mu \text{eV}$. }{The values of $Q_h$, $Q_c$ and $W$ are given in units of $\mu \text{eV}$}.}\label{fig7}
\end{figure}

After the point {where the heat fluxes are inverted that we} discussed previously, the machine starts behaving as a refrigerator, thus we can evaluate the Coefficient of Performance (COP) for it. Similarly as before, we make plots containing all the important information about the refrigerator at this stage. This is carried out in Fig. \ref{fig7} where we put the heat transferred to the hot and to the cold reservoirs, the work done, the COP of a Carnot refrigerator and the COP against the compression ratio $r$. It is important to mention that the definition of COP is only valid in the region III (see Fig. 6). Note that the COP is a monotonically decreasing function of the compression ratio. 

\subsection{The influence of the quantum tunnelling}
Early we have restricted ourselves to the case where there is no change on the tunneling parameters $\Delta_1$ and $\Delta_2$. Although we have found some interesting features for the heat engine, still nothing too different from the classical one was observed. As it is well known, quantum tunneling is not predicted by the laws of classical mechanics: for a particle to surpass a potential barrier it is required potential energy. In the light of recent papers, in particular the Klimovsky work \cite{Klimovsky}, we can extract {some unexpected features} of our machine if we escape from the classical regime, where no work can be done for incompressible working substance, for instance. This can be achieved by varying some gate voltages that control the tunneling parameters of the DQDs.

\begin{figure}[ht!]
	\centering
	\includegraphics[height=5cm,width=6cm]{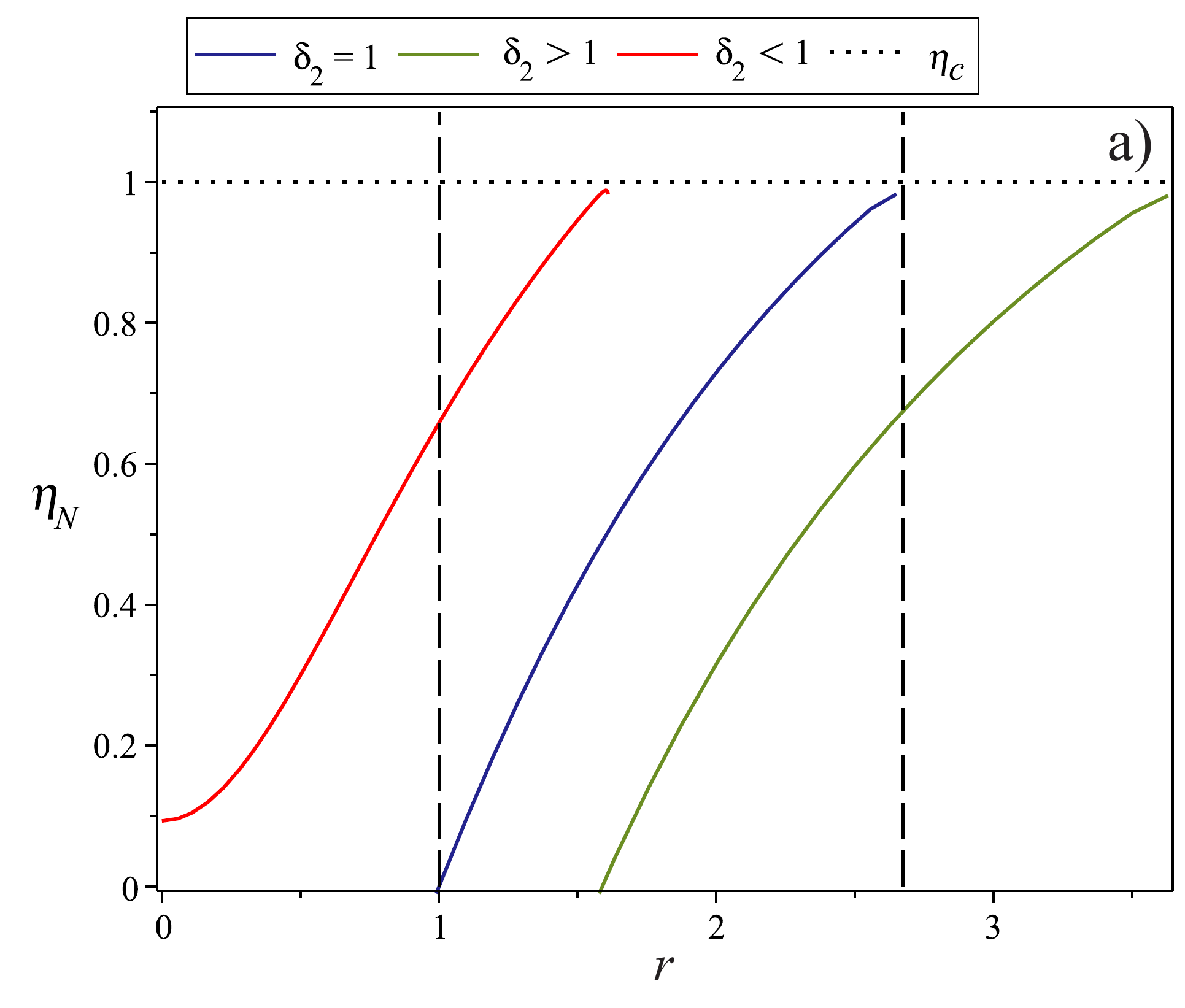}
	\centering
	\includegraphics[height=5cm,width=6cm]{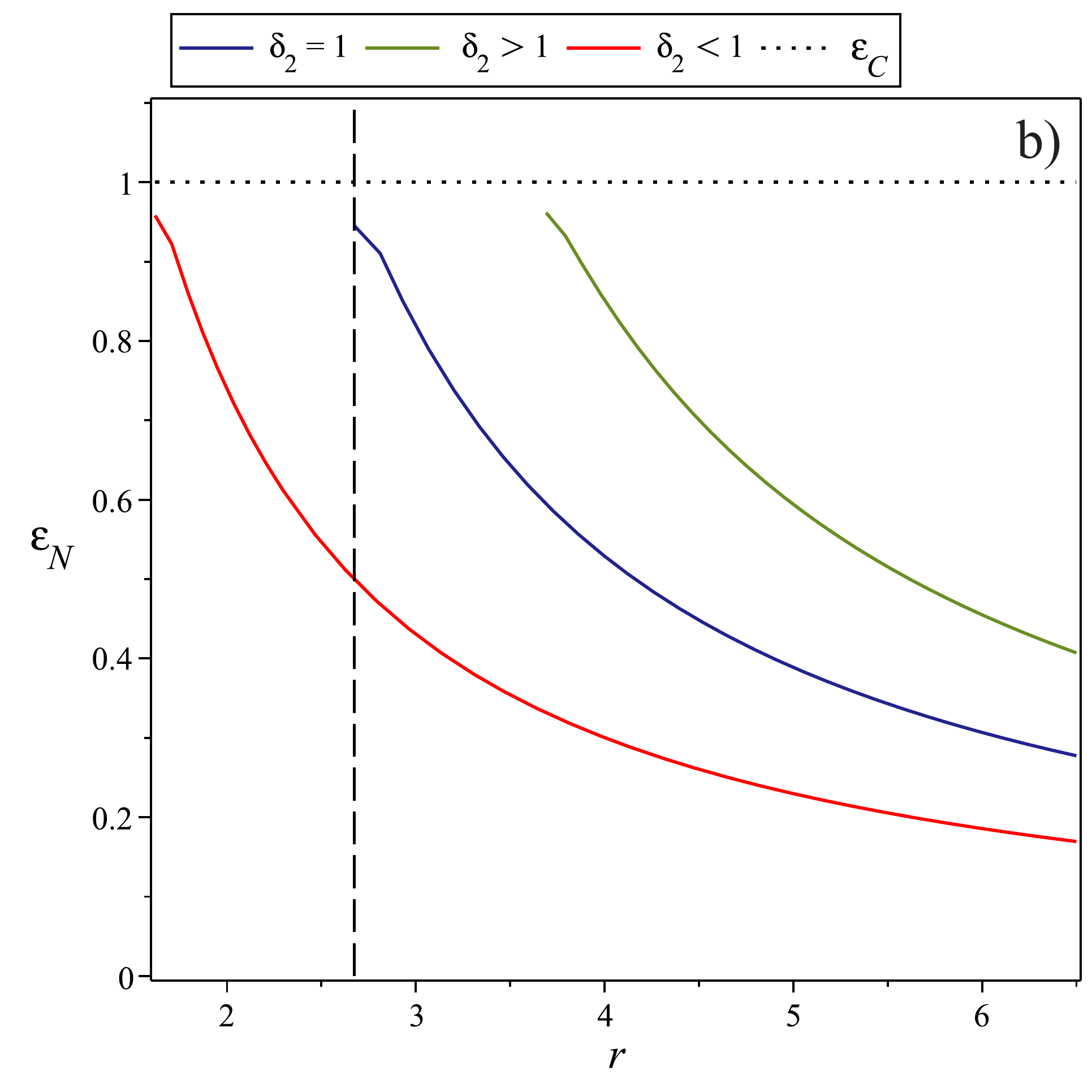}
	\centering
	\includegraphics[height=5.5cm,width=6cm]{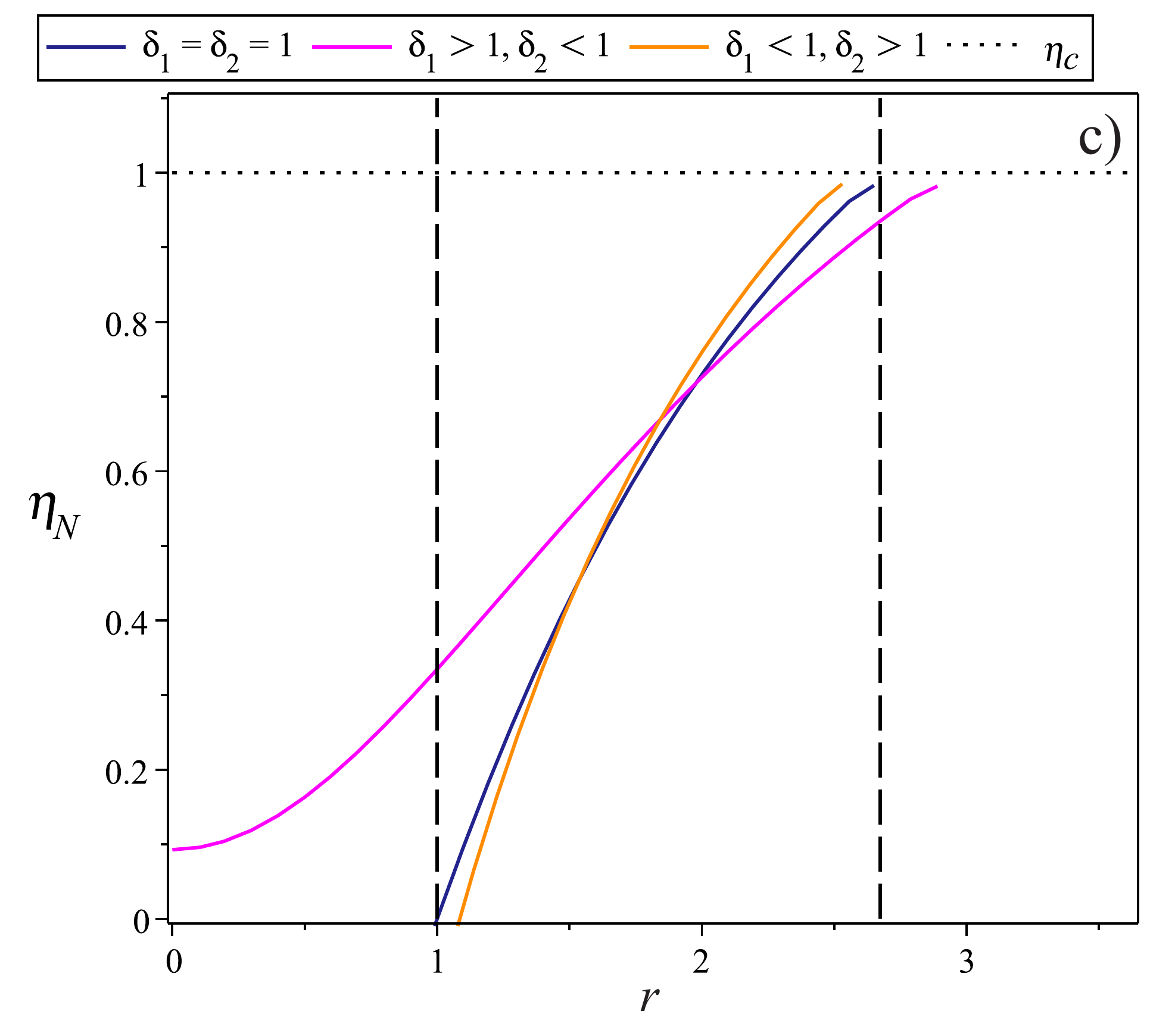}
	\centering
	\includegraphics[height=5.5cm,width=6cm]{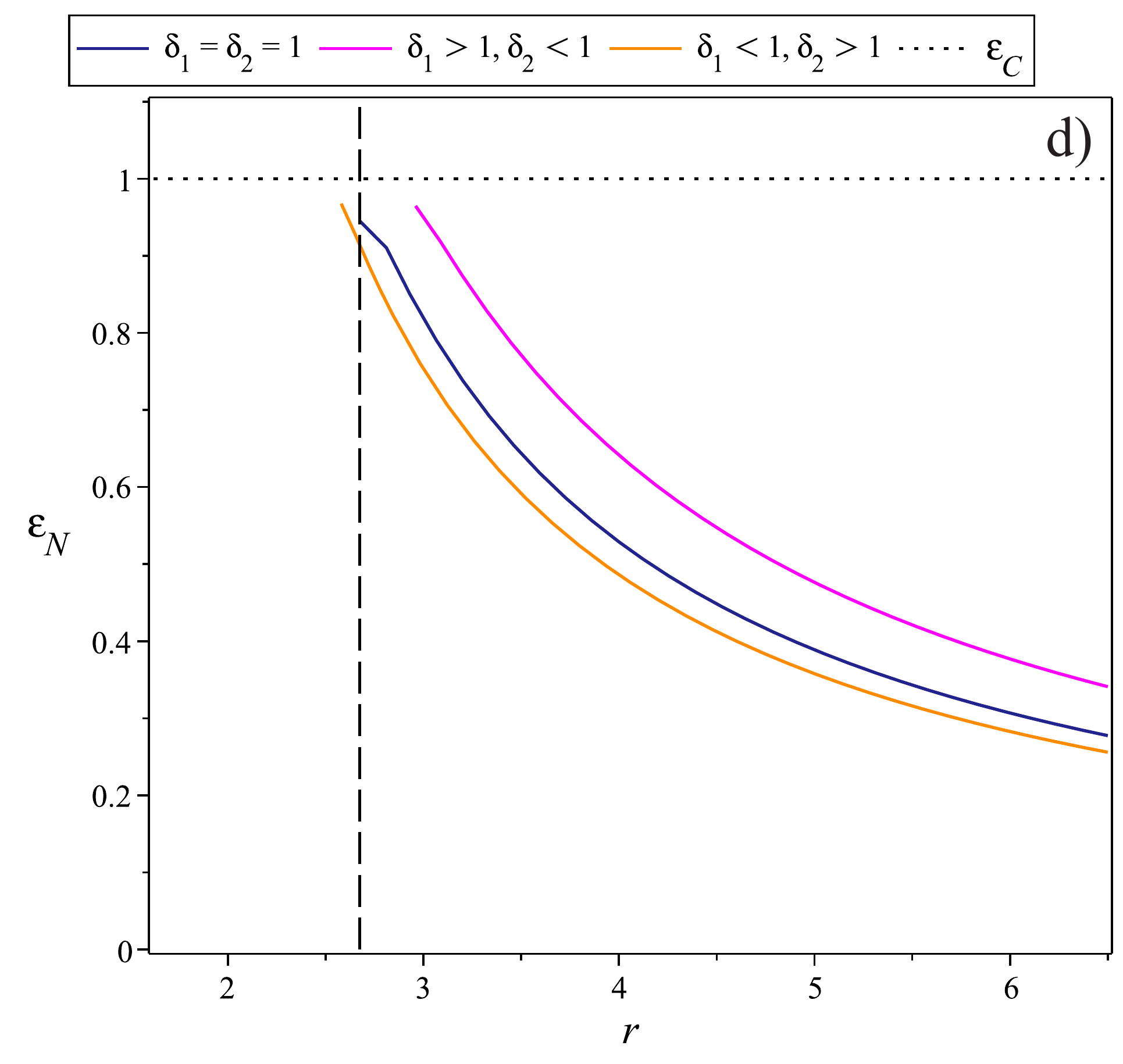}
	\caption{\small In (a), it is depicted the normalized efficiency $\eta_N$ and in (b), it is shown the normalized COP $\varepsilon_N$ curve. It is considered in both of them $\delta_1=1$: the classical case ($\delta_2=1$, blue), the ${\delta_2>1}$ case (with $\Delta_2^c=4\mu \text{eV}$, green) and the ${\delta_2<1}$ case (with $\Delta_2^c=2\mu \text{eV}$, red). In (c) and in (d), we have the normalized efficiency $\eta_N$ and the normalized COP $\varepsilon_N$, respectively: the classical case ($\delta_1=\delta_2=1$) in blue, the stretched magenta curve for $\delta_1>1$ and $\delta_2<1$ (with $\Delta_1^c=18\mu \text{eV}$ and $\Delta_2^c=2\mu \text{eV}$) and the squeezed orange curve for $\delta_1<1$ and $\delta_2>1$ (with $\Delta_1^c=7\mu \text{eV}$ and $\Delta_2^c=4\mu \text{eV}$).
	For all the plots we have set {$V_h=10\mu \text{eV}, \Delta_1^h=10\mu \text{eV},\Delta_2^h=3\mu \text{eV},T_h=2\mu \text{eV},T_c=1\mu \text{eV}$.}}\label{fig8}
\end{figure}

{The classical Otto engine assumes the efficiency $\eta_O=1-\frac{1}{r^{\gamma-1}}$, $\gamma=C_p/C_v$ being the specific heat ratio and $r$ the compression ratio. Note that for $r=1$ the efficiency goes to zero and if $r<1$ the efficiency becomes negative corresponding to the heater regime. This is exactly what happens if we keep $\Delta_1^c=\Delta_1^h$ and $\Delta_2^c=\Delta_2^h$, which is depicted in the blue curves of Fig. \ref{fig8}, thereat we will refer this particular case as the \textit{classical case}. At this point} we can abandon the constraint we made before and consider $\Delta_1^h\neq\Delta_1^c$ and $\Delta_2^h\neq\Delta_2^c$, so that new parameters $\delta_1=\Delta_1^c/\Delta_1^h$ and $\delta_2=\Delta_2^c/\Delta_2^h$ can be defined. In the Fig. \ref{fig8} we plot the efficiency normalized to the Carnot efficiency $\eta_N=\eta/\eta_c$ for some different values of {the quantities} $\delta_1$ and $\delta_2$. We can observe a shift of the curve to the left (right) when we have $\delta_{1(2)}<1$ ($\delta_{1(2)}>1$) individually or we can have a stretching (squeezing) for the left and right if we have $\delta_1>1$ and $\delta_2<1$ ($\delta_1<1$ and $\delta_2>1$) simultaneously, {where $\delta_{1(2)}$ stands for "$\delta_1$ or $\delta_2$".}

In the Fig. \ref{fig8}.(a), {specifically in the red curve corresponding to $\delta_{1}=1$ and $\delta_2<1$}, it is observed an enhancement in the efficiency in comparison to the case where the tunneling parameters are kept fixed throughout the cycle. As a consequence, no heat pump regime appears, which means that the heater was changed to a highly efficient engine and we now have a positive efficiency even for an incompressible working substance, for which $r\equiv1$. Also, the point of inversion from heat engine to refrigerator is shifted to a lower value of $r$. On the other hand, when $\delta_{1(2)}>1$ (green curve), we have a larger region for the operation of the heat pump and the point of inversion from heat engine to refrigerator is also shifted, but to a higher value of $r$ instead. This means that the efficiency is reduced but the region which describes the refrigerator now is valid for highly efficient engine. In Fig. \ref{fig8}.(b), the normalized COP $\varepsilon_N=\varepsilon /\varepsilon_C$, {with $\varepsilon_C$ being the Carnot COP}, is plotted for the same set of parameters values, which reinforces the results previously discussed for the efficiency $\eta$. Moreover, we observe that {the refrigerator presents a better performance} for higher values of $r$ (see the green curve). In the Fig. \ref{fig8}.(c), we can observe a special case for which the efficiency can be enhanced bellow some value of $r$ while it can be diminished above it. This means that the heater is turned into a engine (not to efficient) and the refrigerator is changed to a highly efficient engine (see the magenta curve). The Fig. \ref{fig8}.(d), which stands for the normalized COP $\varepsilon_N=\varepsilon /\varepsilon_C$ completes the analyzes and it shows that a refrigerator can have an improvement in its performance as well (see the magenta curve).

{For almost all of the previous plots, we analyse the compression ratio $r$ increasing up to 6 times, but experimental data \cite{FUJISAWA2011730} shows a variation for interaction coupling $V$ up to almost 3 times at least (the change was from $25 \mu\text{eV}$ to $75\mu\text{eV}$). This way, the previous discussions of the Fig.8 becomes more relevant since we have, in some cases, a shift for the curve to the left, where the compression ratio is lower, supporting the experimental possibility for the realization of all operation modes.}

\begin{figure}[ht!]
		\centering
		\includegraphics[height=7cm,width=8cm]{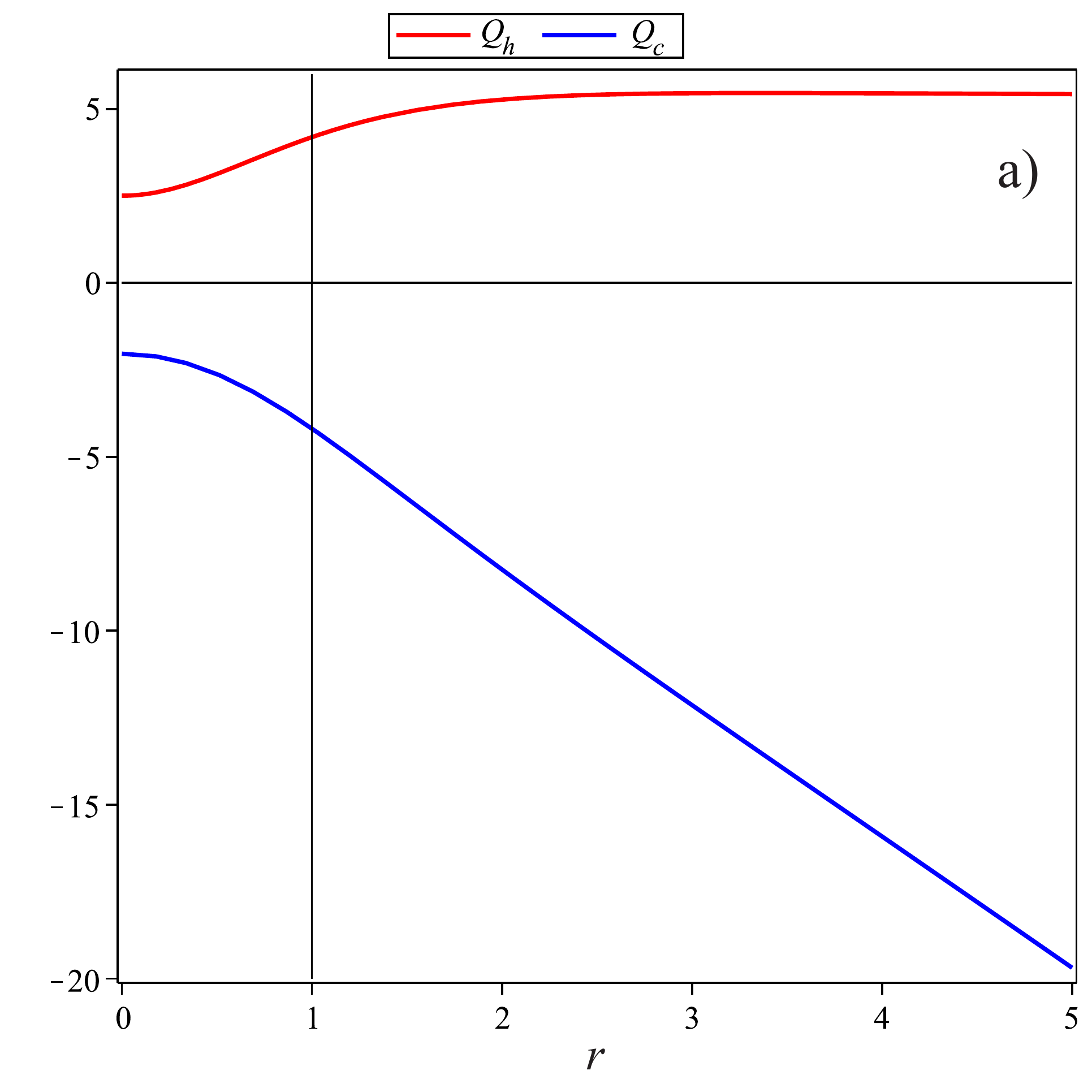}
		\quad
		\centering
		\includegraphics[height=7cm,width=8cm]{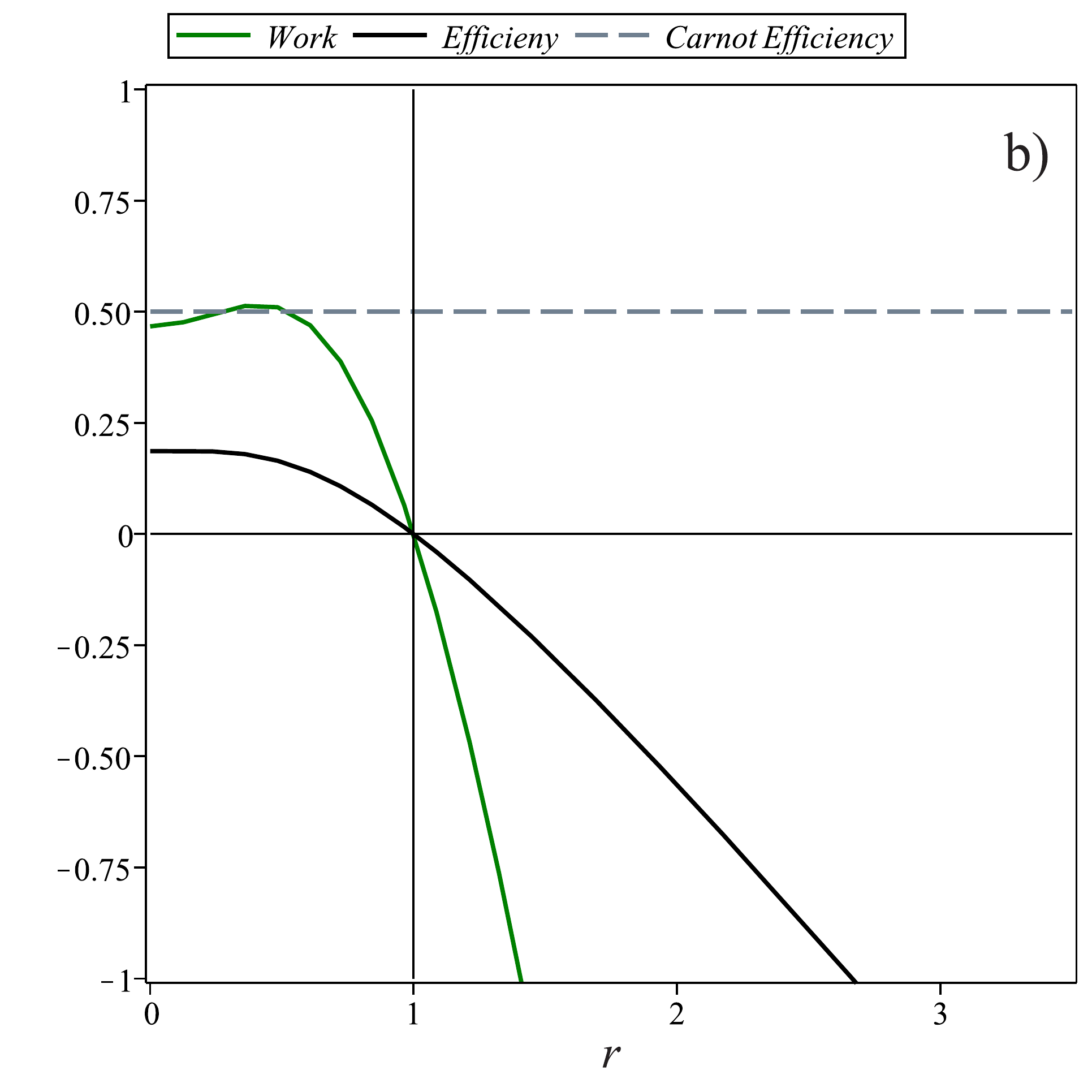}
		\caption{\small In (a) we have the heat exchanged with the hot (red) and cold (blue) reservoirs and in (b) it is depicted the work done (green) and the efficiency (black) against the compression ratio $r$. We have fixed {$V_h=10\mu \text{eV}, \Delta_1^h=10\mu \text{eV},\Delta_2^h=3\mu \text{eV},T_h=20\mu \text{eV},T_c=10\mu \text{eV}$.} {The values of $Q_h$, $Q_c$ and $W$ are given in units of $\mu \text{eV}$}.}\label{fig10}
	\end{figure}

During this whole paper we focus only in {the two-level approach}, but what happens when we have higher temperatures? If we increase the temperatures of the reservoirs the {most excited states in the system becomes relevant (see Appendix B)}. In this case, there is no inversion for the heat exchanged with the reservoirs as we can see in Fig. {\ref{fig10}}.(a). The machine will never turns into a refrigerator, which sustain our explanation from the Fig. {\ref{fig5}} where the inversion of the flow of the heat occurs because of the inversion on the occupation probabilities. Fig. {\ref{fig10}}.(b) shows that the machine {presents very unusual} properties even without varying the tunneling parameters, although no work can be done for incompressible working substance ($r=1$), in the region $r<1$ the machine behaves as a heat engine producing useful work and for $r>1$ the machine will be a heat pump.

Some disadvantages of our model includes a requirement of {approximating our system by a two-level one} for most part of our results and we consider the system completely isolated during the adiabatic processes. It is important to note that the irreversibility of the isochoric processes does not affect the efficiency of the Otto cycle. Our aim for future researches involves considering the finite time cycles, covering the problem in the open quantum systems context \cite{endorev,Thermodynamics}. 
\section{Conclusions}
In conclusion, in this paper, we addressed a theoretical proposal for a quantum heat machine with two sets of coupled DQDs interacting via Coulomb {interaction} of excess electrons inside each DQD, which in turn acts as our charged qubits. We discussed the appearing of different regions of operation for our machine: the heat pump, the heat engine and the refrigerator. These operation modes can be switched by adjusting the {value of the interaction coupling}. We also discussed the reason why these transitions occur and what is truly happening with the machine in the {null work} points. 

Furthermore, we do not just calculate the work done and the efficiency of the heat engine, but also the COP of the refrigerator. In addition, we gave a description on how we can escape the results expected for a classical machine. These findings rely in the effects due to {variations of the parameters that control the} quantum tunnelling of a single electron between each individual DQD. We have observed that the performance of both the engine and the refrigerator can be modified due to {the manipulation of} this well known quantum phenomenon. Also, it allows the modification of the operation mode of the machine: either a heater or a refrigerator can be switched to a highly efficient engine in some cases. We have observed the possibility of work extraction even for an incompressible working substance. Our results follows the spirit of those found in \cite{Klimovsky}, where nonclassical results for a machine were observed but, in their case, as a consequence of a non-homogeneous energy scaling.

{In summary, the present work brings a new example to increment the set of already known quantum heat engines with very promising devices as the working substance, which are the DQDs, also bringing the recipe for the total manipulation of the operation modes. Thereat, the main complications still lies in the coupling with the heat baths and, thinking ahead, how the time can plays a role on this engine, but we leave these questions opened to be explored in future works.}


\section*{Acknowledgements}

This study was financed in part by the Coordenação de Aperfeiçoamento de Pessoal de Nível Superior – Brasil (CAPES) – Finance Code 001. M. Rojas would like to thank CNPq grant 432878/2018-1 and C. Filgueiras would like to thank CNPq grant 305077/2018-0. {We would like to thank the Referees for the valuable suggestions.}

\newpage

\appendix

\section{Work curves for different values of Coulomb {coupling}}

\begin{figure}[ht!]
	\centering
	\includegraphics[height=7cm,width=8cm]{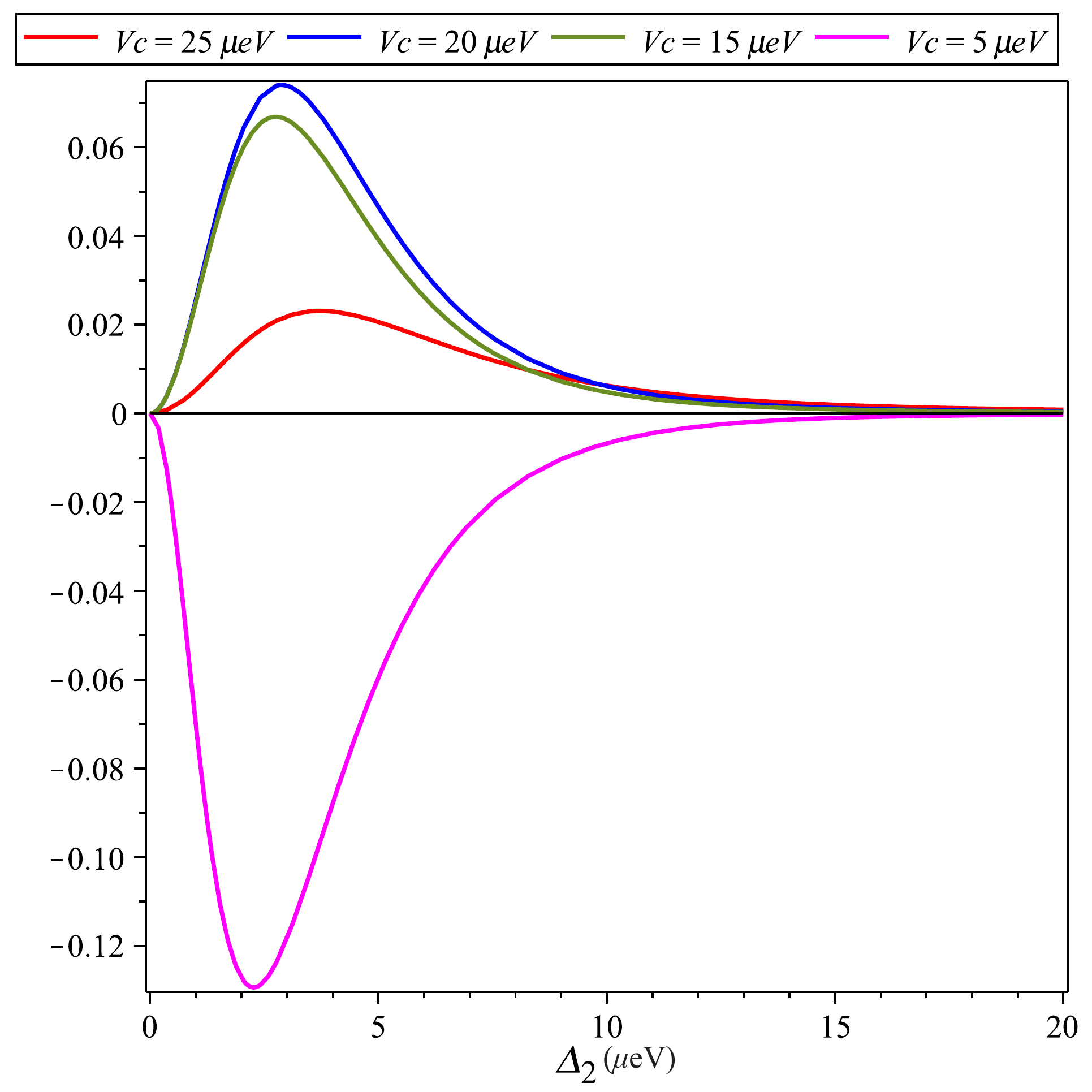}
	\caption{\small The work done against the tunneling parameter $\Delta_2$ for different values of $V_c$ and fixed values: {$V_h=10\mu \text{eV}$ and $\Delta_1=10\mu \text{eV}$. We have $V_c=5\mu \text{eV}$ (magenta curve), $V_c=15\mu \text{eV}$ (green curve), $V_c=20\mu \text{eV}$ (blue curve) and $V_c=25\mu \text{eV}$ (red curve). The case $V_c=10\mu \text{eV}$ is not plotted, but it corresponds to a null work.}}\label{fig9}
\end{figure}

In order to achieve an optimized value for the work done by the heat engine, we plot in the Fig. \ref{fig9} the work done W against the tunneling parameter $\Delta_2$ for some different values of the Coulomb {coupling} $V_c$ (it could also be $\Delta_1$ because, as we discussed early, the Hamiltonian is symmetric). Thus, we fix {$V_h=10\mu \text{eV}$ and $\Delta_1=10\mu \text{eV}$ for the temperatures $T_h=2\mu \text{eV}$ and $T_c=1\mu \text{eV}$}, and we can see that the peak of the curve does not shift considerable, so we can approximately estimate the best value for $\Delta_2$, at least for the order of magnitude, that optimize the work done, to know {$\Delta_2\approx3\mu \text{eV}$}.

Note that, in concordance with Fig. \ref{fig4}, increasing the {interaction coupling} $V_c$ will increase the work done for a while (green and blue curves), and then it will start to decrease it (red curve). Also we see that for $V_c<V_h$, the work becomes drastically negative (yellow curve), which also agrees with the positive work condition stated in the Fig. \ref{fig4}.

\newpage

\section{Probabilities distribution}

\begin{figure}[ht!]
	\centering
	\includegraphics[height=7cm,width=8cm]{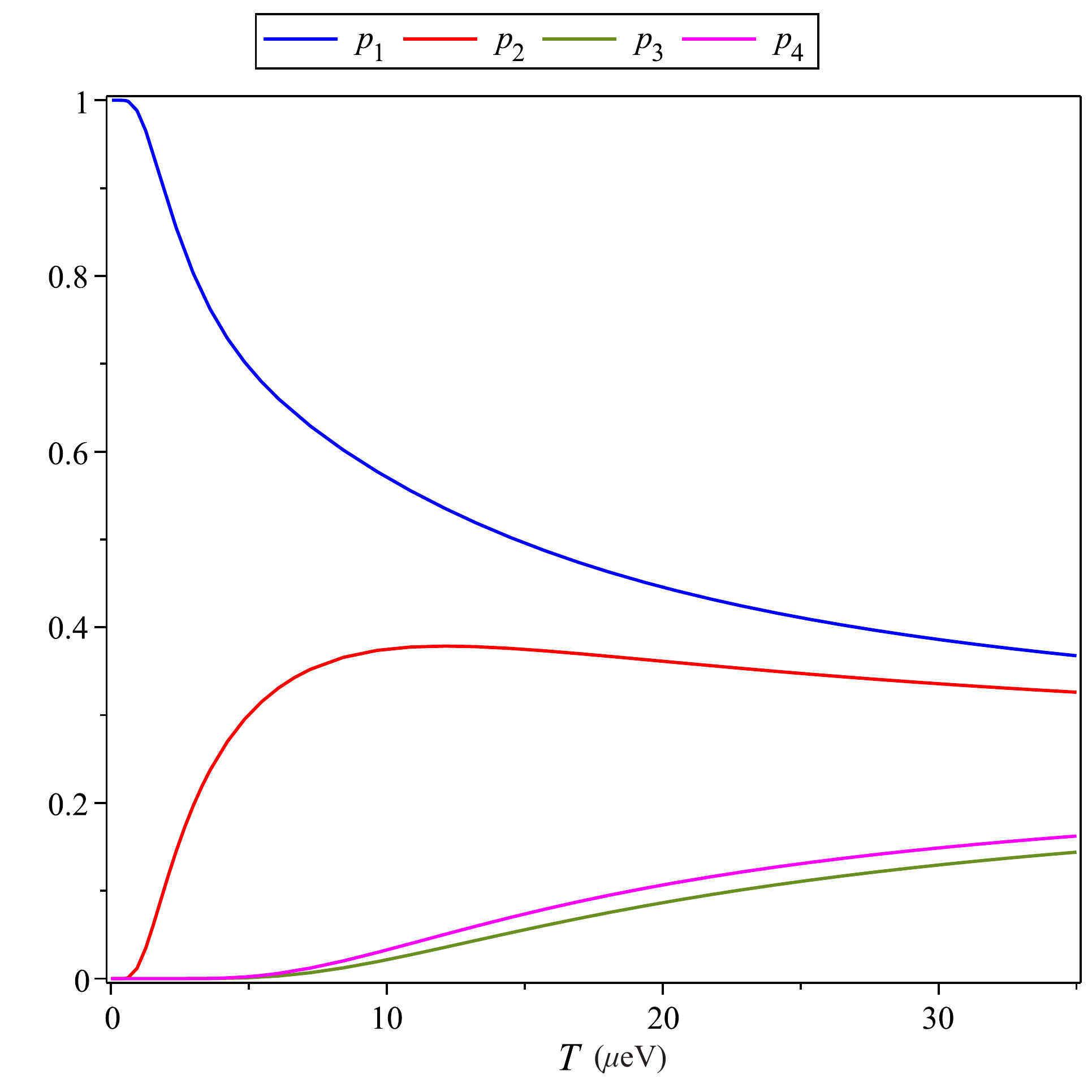}
	\caption{\small {The occupation probabilities $p_n$ against the temperature $T$ for fixed values: $V=10\mu \text{eV}$, $\Delta_1=10\mu \text{eV}$ and $\Delta_2=3\mu \text{eV}$.}}\label{fig11}
\end{figure}

{In the Fig. \ref{fig11} we see the occupation probabilities distribution $p_n=exp(-E_n/kT)/Z$ for the four possible states $\ket{\psi_n}_{n=1..4}$ in terms of the temperature. We see that for $T=1\mu \text{eV}$ or even $T=2\mu \text{eV}$ (values used in most of our plots), the second and the third excited states has practically zero probability ($p_3\approx p_4 \approx 0$) for the system to be found in, this way we can neglect this two most excited states and approximate our system by a two level one. On the other hand, if we have $T=10\mu \text{eV}$ or $T=20\mu \text{eV}$ (values used in Fig. \ref{fig10}), we see that $p_3$ and $p_4$ are no longer negligible, thus the two-level approximation is no longer valid.}

\newpage




\bibliographystyle{elsarticle-num-names}
\bibliography{sample.bib}







\end{document}